\documentclass[prd,onecolumn,notitlepage,nofootinbib,superscriptaddress,showpacs,showkeys]{revtex4-2}
\usepackage{amsfonts}
\usepackage{graphicx}
\usepackage{dcolumn}
\usepackage{amsmath}
\usepackage{amssymb}
\usepackage[T1]{fontenc}
\usepackage[utf8]{inputenc}
\usepackage[brazil,english]{babel}
\usepackage[usenames,dvipsnames]{color}
\usepackage{longtable}
\usepackage{ulem}
\usepackage{bm}
\usepackage{graphicx}
\usepackage{cancel}

\begin{document}
	
	\title{Quasi-local Hamiltonian for generalized Kerr-Schild black holes in the Iyer-Wald formalism}
	\author{M. \thinspace A. Jaraba}
	\email{markjarava@gmail.com}
	\affiliation{Universidade Estadual de Londrina, Departamento de F\'{\i}sica, CEP 86051-990, Londrina-PR, Brazil}
	\author{T.\thinspace L. Campos}
	\email{thiagocampos@alumni.usp.br}
	\affiliation{Centro Brasileiro de Pesquisas Físicas, CEP 22290-180, Rio de Janeiro-RJ, Brazil}
	\author{M.\thinspace C. Baldiotti}
	\email{baldiotti@uel.br}
	\affiliation{Universidade Estadual de Londrina, Departamento de F\'{\i}sica, CEP 86051-990, Londrina-PR, Brazil}
	
	\begin{abstract}
		Grounded in the covariant phase space approach of the Iyer-Wald formalism, we propose a quasi-local Hamiltonian for generalized Kerr-Schild (GKS) black holes. Evaluated for spherically symmetric spacetimes over a flat Minkowski background, our formulation recovers the Misner-Sharp mass. Because this framework relies solely on the GKS structure, it provides a straightforward extension of the Misner-Sharp mass to rotating geometries. We further extend our analysis to asymptotically anti-de Sitter (AdS) black holes, where a renormalization procedure to subtract the background AdS energy naturally emerges. By employing this technique, we establish a geometric quasi-local formulation of Kerr-AdS thermodynamics.
	\end{abstract}
	
	\maketitle
	
	\section{Introduction}
	
	The concept of gravitational energy in general relativity is notoriously subtle. The rank-2 nature of the energy-momentum tensor, combined with spacetime curvature, hinders the formulation of local currents that yield globally conserved charges in generic geometries, particularly in the absence of stationary Killing fields \cite{penrose1982quasi,penrose1982some}. Furthermore, due to the Equivalence Principle, a local, covariant energy density for the gravitational field does not exist \cite{17,4}. Consequently, covariant energy definitions are inherently global or quasi-local, and can be broadly classified into two groups: non-canonical (purely geometric) and canonical (based on the action).
	
	Canonical approaches have been developed through three main formulations: superpotential methods, asymptotic Hamiltonian constructions, and the quasi-local Brown-York definition. Superpotential methods rely on evaluating conserved charges associated with diffeomorphisms. The standard case uses the Einstein-Hilbert Lagrangian. However, the Noether charge obtained directly from this action yields only half of the physical mass for stationary black holes in asymptotically flat spacetimes. To correct this and reproduce the proper mass, an \textit{ad hoc} factor of $2$ is required \cite{carroll}, leading to the well-known Komar integral \cite{16}. Attempts to regularize this method introduced additional structures, such as the bimetric Lagrangian proposed by Katz \cite{katz} and generalized in the KBL formalism \cite{KBL}.
	
	An alternative to superpotentials is the Hamiltonian formulation, based on the ADM foliation \cite{21}. To obtain a well-defined variational principle in phase space from a totally constrained Hamiltonian, Regge and Teitelboim \cite{20} demonstrated the necessity of adding asymptotic boundary terms to this Hamiltonian. Consequently, these methods provide global definitions of energy evaluated at spatial infinity (ADM mass) or null infinity (Bondi-Sachs mass \cite{22, hollanbondi, 23, ashtekarbondi}), which can be extended to asymptotically anti-de Sitter (AdS) spacetimes \cite{teitelboim}.
	
	Finally, to bypass the asymptotic limits required by traditional Hamiltonian methods, Brown and York introduced a quasi-local approach based on the Hamilton-Jacobi formalism applied to the gravitational action \cite{18} with the Gibbons-Hawking-York boundary term \cite{hawkinghoro}. Although this formulation allows for the calculation of conserved quantities on finite spatial 2-surfaces, its energy diverges even in the flat space limit. To address this, the original Brown-York regularization requires an isometric embedding into a reference background to set the zero of energy. Because this embedding can be challenging in AdS spacetimes, holographic renormalization was introduced to provide boundary counterterms.
	
	In the non-canonical category, the Misner-Sharp mass is notable for its operational simplicity. Originally formulated for spherically symmetric spacetimes using geometric invariants, it has proven highly effective in the context of black-hole mechanics and thermodynamics. Hayward analyzed the behavior of the Misner-Sharp mass across different classes of trapping horizons, demonstrating that it satisfies a unified first law of horizon dynamics \cite{hayward1994general,hayward1996gravitational,hayward1998unified}. It was subsequently shown that this mass is associated with a dynamical Hawking temperature derived via the tunneling method of Parikh and Wilczek \cite{hayward2009local, parikh2000hawking}. Furthermore, Padmanabhan \cite{padmanabhan2002classical} revealed a profound equivalence between Einstein's field equations evaluated at the horizon and fundamental thermodynamic identities, a context in which the Misner-Sharp mass plays the role of the thermodynamic energy of the system.
	
	The canonical definitions cited above find a unified framework within the Covariant Phase Space formalism. Introduced by Lee, Wald, and Iyer \cite{1,2,3,iyer1995comparison}, this construction generalizes the notion of the Hamiltonian to diffeomorphism-invariant theories based on Lagrangians that can depend on higher-order derivatives of the fields. The variation of the Hamiltonian is then constructed from the symplectic current, a $(D-1)$-form derived locally from the Lagrangian in a $D$-dimensional spacetime. As Wald and Iyer demonstrated \cite{2,3}, the first law of black-hole thermodynamics can be derived directly from the Lagrangian, defining entropy as a Noether charge (Wald entropy). This approach, the Iyer-Wald formalism, is valid for any gravity theory based on a diffeomorphism-invariant Lagrangian.
	
	The Iyer-Wald formalism provides a robust framework for deriving Hamiltonian variations associated with spacetime symmetries. Obtaining the corresponding Hamiltonians requires the Hamiltonian variation to be integrable in phase space. While this variation is initially defined as a bulk integral over a Cauchy surface, this formalism allows it to be recast as a boundary integral, providing a quasi-local description. To our knowledge, the application of this formalism to finite domains has not been thoroughly explored in the literature. Therefore, the main objective of this paper is to investigate this exact quasi-local formulation by providing a prescription for obtaining the Hamiltonian of generalized Kerr-Schild (GKS) black holes. Specifically, we propose a Hamiltonian 2-form, derived purely from the gravitational sector, to calculate the quasi-local energy within a finite region of spacetime. This approach eliminates the need to add counterterms to the action or to impose asymptotic boundary conditions. As a relevant particular case, we demonstrate that the Misner-Sharp mass, despite its non-canonical (geometric) origin, can be derived as a quasi-local Hamiltonian charge within this framework. Consequently, since our construction relies only on the GKS form of the metric, it allows us to propose a generalization of the Misner-Sharp mass for rotating geometries. Furthermore, we show that even in an AdS background our construction provides a consistent quasi-local perspective on the traditional thermodynamics of Kerr-AdS black holes.
	
	The paper is organized as follows. In Section~\ref{sec:iyer_wald}, we review the covariant phase space approach and the Iyer-Wald formalism. Building upon these foundations, Section~\ref{sec:quasi_local_hamiltonian} introduces our proposed quasi-local Hamiltonian for GKS metrics. In Section~\ref{sec:applicationsMin}, we apply this formulation to black holes in a Minkowski background, demonstrating that, in the spherically symmetric case, it agrees with the Misner-Sharp mass, which can then be naturally extended to rotating spacetimes. Section~\ref{sec:applicationsAdS} analyzes asymptotically AdS black holes, where we introduce a renormalization technique to subtract the background AdS energy, thereby allowing a direct connection to Kerr-AdS thermodynamics. Final remarks are outlined in Section~\ref{remarks}. We use the geometric unit system and signature $(-,+,+,+)$ for the metric.
	
	\section{Covariant Phase Space and Iyer-Wald Formalism}
	\label{sec:iyer_wald}
	
	The covariant phase space formalism, developed by Lee, Wald, and Iyer \cite{1,2,3}, provides a framework to construct conserved quantities and the symplectic structure of a theory directly from its Lagrangian. Consider a theory of dynamical tensor fields $\phi=\{g_{\mu\nu},\Phi\}$, where $g_{\mu\nu}$ is the metric and $\Phi$ represents all other dynamical fields, on a $D$-dimensional spacetime, governed by a diffeomorphism-invariant Lagrangian $D$-form $\mathbf{L}(\phi)$. A general first-order variation of the Lagrangian with respect to the dynamical fields yields 
	\begin{equation}
		\delta\mathbf{L}=\mathbf{E}(\phi)\delta\phi+\mathrm{d}\boldsymbol{\Theta}(\phi,\delta\phi)~, \label{deflang}
	\end{equation}
	where $\mathbf{E}(\phi)=0$ represents the classical equations of motion and $\boldsymbol{\Theta}$ is the symplectic potential, a $(D-1)$-form constructed from the fields and their variations, linear in $\delta\phi$ and its derivatives.
	
	Associated with any smooth vector field $\xi$ generating a one-parameter group of diffeomorphisms, one defines the Noether current $(D-1)$-form as 
	\begin{equation}
		\mathbf{J}[\xi]=\boldsymbol{\Theta}(\phi,\mathcal{L}_{\xi}\phi)-\iota_{\xi}\mathbf{L}~,
		\label{Jcorrent}
	\end{equation}
	where $\mathcal{L}_{\xi}$ denotes the Lie derivative along $\xi$ and $\iota_{\xi}$ is the interior product. Using the Cartan identity and the diffeomorphism-invariance of the Lagrangian, the exterior derivative of the current is given by $\mathrm{d}\mathbf{J}[\xi]=-\mathbf{E}\mathcal{L}_{\xi}\phi$ \cite{1}. On-shell ($\mathbf{E}=0$), the current is closed and can be locally expressed as the exact differential of a $(D-2)$-form, defining the Noether charge $\mathbf{Q}[\xi]$, 
	\begin{equation}
		\mathbf{J}[\xi] =\mathrm{d}\mathbf{Q}[\xi]~.
		\label{NoetherCharge}
	\end{equation}
	The symplectic structure of the theory is encoded in the symplectic current $(D-1)$-form, defined by taking an antisymmetrized variation of the symplectic potential \cite{3}: 
	\begin{equation}
		\boldsymbol{\omega} ( \phi,\delta_1\phi,\delta_2\phi ) =\delta_1\boldsymbol{\Theta}(\phi,\delta_2\phi)-\delta_2\boldsymbol{\Theta}(\phi,\delta_1\phi)~.
	\end{equation}
	The integration of $\boldsymbol{\omega}$ over an achronal Cauchy surface $\Sigma$ defines the pre-symplectic form 
	\begin{equation}
		\Omega(\phi,\delta_1\phi,\delta_2\phi)=\int_{\Sigma}\boldsymbol{\omega}~. \label{FormaSim}
	\end{equation}
	Provided that suitable boundary conditions are imposed on the dynamical fields at spatial infinity, $\Omega$ becomes independent of the choice of $\Sigma$ \cite{3,23}. Factoring out the degenerate directions of $\Omega$ leads to the physical phase space of the theory.
	
	If $\xi$ generates a phase space evolution, the variation of the Hamiltonian $H_{\xi}$ associated with this flow is completely determined by the pre-symplectic form \cite{3,23}: 
	\begin{equation}
		\delta H_{\xi} =\Omega(\phi,\delta\phi, \mathcal{L}_{\xi}\phi)~. \label{DefdeHamilto}
	\end{equation}
	This Hamiltonian variation can be directly related to the Noether charge. By varying the Noether current \eqref{Jcorrent} and keeping the vector field $\xi$ fixed, one finds that, on-shell \cite{3}, 
	\begin{equation}
		\boldsymbol{\omega}(\phi,\delta\phi,\mathcal{L}_{\xi}\phi)=\delta\mathrm{d} \boldsymbol{Q}[\xi] -\mathrm{d}(\iota_{\xi}\boldsymbol{\Theta})~. \label{EQVART}
	\end{equation}
	Integrating this relation over $\Sigma$ and applying Stokes' theorem yields 
	\begin{equation}
		\delta H_{\xi}= \int_{\partial\Sigma}\left(\delta\mathbf{Q}[\xi]-\iota_{\xi}\boldsymbol{\Theta}(\phi,\delta\phi)\right)~, \label{Hbor}
	\end{equation}
	where $\partial\Sigma$ is the boundary of $\Sigma$. Thus, the Hamiltonian in a diffeomorphically covariant theory is formulated as a boundary integral.
	
	For spacetimes admitting asymptotically flat solutions, one can construct global conserved quantities associated with asymptotic symmetries. If $\xi$ approaches a Killing vector field of the asymptotic Minkowski background (i.e., $\mathcal{L}_{\xi}g_{\mu\nu}$ satisfies appropriate asymptotic conditions \cite{5}), the Hamiltonian is integrable provided there exists a $(D-1)$-form $\mathbf{B}$ such that the pullback to $\partial\Sigma$ satisfies 
	\begin{equation}
		\delta\int_{\infty} \iota_{\xi}\mathbf{B}=\int_{\infty}\iota_{\xi}\boldsymbol{\Theta}~. \label{arb}
	\end{equation}
	Under this integrability condition, global charges such as energy $\mathcal{E}$ and angular momentum $\mathcal{J}$ can be defined by associating $\xi$ with asymptotic time translations ($t$) and axial rotations ($\varphi$), respectively \cite{2,3}: 
	\begin{equation}
		\mathcal{E}= \int_{\infty}\left(\mathbf{Q}[t]-\iota_t\mathbf{B}\right)~,\qquad\quad\mathcal{J}=-\int_{\infty}\mathbf{Q}[\varphi]~. \label{energiayangulo}
	\end{equation}
	These definitions have been shown to match the standard Arnowitt-Deser-Misner (ADM) global charges \cite{3}.
	
	\section{Quasi-local Hamiltonian for generalized Kerr-Schild metrics}
	\label{sec:quasi_local_hamiltonian}
	
	In this section, we explore the possibility of defining a Hamiltonian from a quasi-local perspective for GKS spacetimes. Suppose that one can find a $(D-2)$-form $\boldsymbol{\mathcal{B}}(\phi)$ such that 
	\begin{equation}
		-\iota_{\xi} \boldsymbol{\Theta}(\phi,\bar{\delta}\phi)=\bar{\delta} \boldsymbol{\mathcal{B}}(\phi) \label{Eqcomp}
	\end{equation}
	is satisfied over some restricted subset $\mathcal{F}$ of the configuration space. Let $\phi_0\in\mathcal{F}$ be a solution to the equations of motion, with $\bar{\delta}$ representing the variation restricted to that domain. Then Eq.~\eqref{EQVART} reduces to 
	\begin{equation}
		\bar{\delta} \boldsymbol{\mathcal{H}}=\left.\boldsymbol{\omega}(\phi,\bar{\delta}\phi,\mathcal{L}_{\xi}\phi)\right\vert_{\phi=\phi_0}~, \label{DeflocalHaml}
	\end{equation}
	where $\boldsymbol{\mathcal{H}}=\mathrm{d}\mathbf{H}$, and the $(D-2)$-form $\mathbf{H}$ is defined by 
	\begin{equation}
		\mathbf{H}=\mathbf{Q}+\boldsymbol{\mathcal{B}}~.
	\end{equation}
	If such a $\boldsymbol{\mathcal{B}}$ exists, we interpret Eq.~\eqref{DeflocalHaml} as a quasi-local version of Eq.~\eqref{DefdeHamilto} over the domain $\mathcal{F}$, where $\boldsymbol{\mathcal{H}}$ represents the Hamiltonian current, and $\mathbf{H}$ is the quasi-local Hamiltonian $(D-2)$-form for the solution $\phi_0$.
	
	Let us consider the case of general relativity with a nonvanishing cosmological constant $\Lambda$ in dimension $D=4$. The action functional for general relativity contains contributions from the gravitational and matter fields. The Lagrangian for the bulk is thus $\boldsymbol{L}=\boldsymbol{L}^\text{EH}+\boldsymbol{L}^\text{matter},$ in which the Einstein-Hilbert Lagrangian $4$-form is
	\begin{equation}
		\mathbf{L}^{\text{EH}}=\frac{1}{16\pi}(R-2\Lambda)\boldsymbol{\epsilon}~, \label{LANGREL}
	\end{equation}
	where $\boldsymbol{\epsilon}$ is the volume form and $R$ is the Ricci scalar. In this work, the quasi-local charges will be constructed purely from the gravitational (geometric) sector. Thus, although our construction is not restricted to vacuum solutions, the matter Lagrangian does not contribute directly to these charges. Instead, the effect of the matter fields will be incorporated solely through the corresponding metric solutions of the Einstein equations. 
	
	Within this framework, the components of the corresponding symplectic potential boundary term $\boldsymbol{\Theta}$ are given by 
	\begin{equation}
		\Theta_{\nu\alpha\beta}= -\frac{1}{16\pi}\epsilon_{\mu\nu\alpha\beta}\left[\nabla_\sigma\delta g^{\mu\sigma}-\nabla^\mu(g^{\sigma\rho}\delta g_{\sigma\rho})\right]~. \label{ThetaGR}
	\end{equation}
	From $\boldsymbol{\Theta}$ and $\mathbf{L}^{\text{EH}}$, we construct the Noether current associated with the vector field $\xi$, 
	\begin{equation}
		J_{\nu\alpha\beta} =\frac{1}{8\pi}\epsilon_{\mu\nu\alpha\beta}\nabla_\sigma\left(\nabla^{[\sigma}\xi^{\mu]}\right)+\frac{1}{8\pi}\epsilon_{\mu\nu\alpha\beta}\left[R_\sigma^\mu+\left(\Lambda-\frac{R}{2}\right)\delta_\sigma^\mu\right]\xi^\sigma~,
	\end{equation}
	leading to its respective Noether charge, 
	\begin{equation}
		Q_{\mu\nu} = -\frac{1}{16\pi}\epsilon_{\mu\nu\alpha\beta}\nabla^\alpha\xi^\beta~. \label{kommarmass}
	\end{equation}
	
	Using $\boldsymbol{\Theta}$ given by Eq.~\eqref{ThetaGR}, we will show that it is possible to find a $\boldsymbol{\mathcal{B}}$ compatible with Eq.~\eqref{Eqcomp} if we choose $\mathcal{F}$ as the set of metrics admitting the GKS form, i.e., metrics written as
	\begin{equation}
		g_{\mu\nu}=\bar{g}_{\mu\nu}+fk_\mu k_\nu~, \label{KSForm}
	\end{equation}
	where $\bar{g}_{\mu\nu}$ is an arbitrary but fixed background metric, while the scalar field $f$ and the null vector field $k^\mu$ are treated as dynamical variables.
	
	Spacetimes in the GKS form possess well-known geometric properties \cite{15}: 
	\begin{align}
		&g^{\mu\nu}=\bar{g}^{\mu\nu}-f(x)k^\mu k^\nu~,\qquad \det(g_{\mu\nu})=\det(\bar{g}_{\mu\nu})~, \qquad \Gamma_{\alpha\beta}^\mu k^\alpha k^\beta=\bar{\Gamma}_{\alpha\beta}^\mu k^\alpha k^\beta~, \notag \\
		& \Gamma_{\alpha\beta}^\mu k_\mu=\bar{\Gamma}_{\alpha\beta}^\mu k_\mu-\frac{1}{2}k^\lambda\nabla_\lambda(fk_\alpha k_\beta)~,\qquad \Gamma_{\mu\beta}^\alpha k^\mu=\bar{\Gamma}_{\mu\beta}^\alpha k^\mu+\frac{1}{2}k^\lambda\nabla_\lambda(fk^\alpha k_\beta)~, \label{Eqsimitricprop}
	\end{align}
	in which $k^\mu=g^{\mu\nu}k_\nu=\bar{g}^{\mu\nu}k_\nu$, and $\Gamma$ and $\bar{\Gamma}$ are the Christoffel symbols associated with $g_{\mu\nu}$ and $\bar{g}_{\mu\nu}$, respectively. Using these properties, it follows that 
	\begin{equation}
		\nabla_\sigma k^\sigma=\bar{\nabla}_\sigma k^\sigma~,\qquad k^\mu\nabla_\mu k^\nu=k^\mu\bar{\nabla}_\mu k^\nu~,\qquad \nabla_\alpha k^\beta=\bar{\nabla}_\alpha k^\beta+\frac{1}{2}k^\lambda\nabla_\lambda(fk^\alpha k_\beta)~, \label{Eqsimten}
	\end{equation}
	where $\bar{\nabla}$ is the connection compatible with $\bar{g}_{\mu\nu}$.
	
	Assuming a metric in the GKS form, the restricted variation of the metric in the subset $\mathcal{F}$ is given by 
	\begin{equation}
		\bar{\delta}g_{\mu\nu}=fk_\mu\delta k_\nu+fk_\nu\delta k_\mu+k_\mu k_\nu\delta f~,\qquad \bar{\delta}g^{\mu\nu}=-fk^\mu\delta k^\nu-fk^\nu\delta k^\mu-k^\mu k^\nu\delta f~. \label{varf}
	\end{equation}
	Substituting Eqs.~\eqref{KSForm} and \eqref{varf} into Eq.~\eqref{ThetaGR}, we obtain 
	\begin{align}
		\left(-\iota_\xi\boldsymbol{\Theta}\right)_{\alpha\beta}&=\frac{1}{16\pi}\epsilon_{\alpha\beta\mu\nu}\xi^\nu\left[\nabla_\sigma(\bar{\delta}g^{\sigma\mu})-\nabla^\mu(g^{\sigma\rho}\bar{\delta}g_{\sigma\rho})\right] \notag \\
		&=-\frac{1}{16\pi}\epsilon_{\alpha\beta\mu\nu}\xi^\nu\left[\bar{\delta}\bar{\nabla}_\sigma(k^\sigma k^\mu f)+f^2 k^\mu k^\sigma\bar{\nabla}_\sigma k^\lambda\delta k_\lambda\right]~, \label{hjl}
	\end{align}
	where we have explicitly used the null condition $k_\mu k^\mu=0$ and the properties in Eq.~\eqref{Eqsimitricprop}. If we further assume a solution where $k^\mu$ is geodesic, 
	\begin{equation}
		k^\sigma\nabla_\sigma k^\lambda=k^\sigma\bar{\nabla}_\sigma k^\lambda=\alpha(x)k^\lambda~,
	\end{equation}
	for some scalar function $\alpha$, the second term in the last line of Eq.~\eqref{hjl} vanishes (since $k^\lambda\delta k_\lambda=0$). From the full metric compatibility $\nabla_\sigma g^{\sigma\mu}=0$, which combined with the KS metric yields $\nabla_\sigma(fk^\sigma k^\mu)=\nabla_\sigma\bar{g}^{\sigma\mu}$, also noting that the background metric is fixed under $\bar{\delta}$ and $\boldsymbol{\epsilon}=\bar{\boldsymbol{\epsilon}}$, we can write 
	\begin{equation}
		\left(-\iota_\xi\boldsymbol{\Theta}\right)_{\alpha\beta}=\left.\bar{\delta}\left[-\frac{1}{16\pi}\epsilon_{\alpha\beta\mu\nu}\xi^\nu\nabla_\sigma\bar{g}^{\sigma\mu}\right]\right\vert_{\text{KS geodesic}}~.
	\end{equation}
	Thus, a 2-form $\boldsymbol{\mathcal{B}}$ compatible with Eq.~\eqref{Eqcomp} is given by 
	\begin{equation}
		B_{\mu\nu}=-\frac{1}{16\pi}\epsilon_{\mu\nu\alpha\beta}\nabla_\sigma\bar{g}^{\sigma\alpha}\xi^\beta+C_{\mu\nu}(\bar{g}_{\alpha\beta})~,
	\end{equation}
	where $C_{\mu\nu}$ is a 2-form locally constructed from the background metric $\bar{g}_{\mu\nu}$.
	
	From this construction, we conclude that for solutions of the Einstein field equations with a cosmological constant that admit a GKS form with a geodesic $k^\mu$, we can define the following Hamiltonian 2-form: 
	\begin{equation}
		H_{\mu\nu}(\xi)=-\frac{1}{16\pi}\epsilon_{\mu\nu\alpha\beta}\left(\nabla^\alpha\xi^\beta+\nabla_\sigma\bar{g}^{\sigma\alpha}\xi^\beta\right)+C_{\mu\nu}(\bar{g}_{\alpha\beta})~. \label{energylocal}
	\end{equation}
	
	The quasi-local Hamiltonian derived in Eq.~\eqref{energylocal} contains the 2-form $C_{\mu\nu}(\bar{g}_{\alpha\beta})$ constructed entirely from the background metric. Hence, we identify this term as the reference Hamiltonian of the background spacetime itself. Consequently, the physically relevant quantity characterizing the black-hole perturbation is the difference $\Delta_{\bar{g}}H_{\mu\nu}\equiv H_{\mu\nu}-C_{\mu\nu}$. Thus, we define the quantity 
	\begin{equation}
		\Delta_{\bar{g}}H_{\mu\nu}(\xi)\equiv-\frac{1}{16\pi}\epsilon_{\mu\nu\alpha\beta}\left(\nabla^\alpha\xi^\beta+\nabla_\sigma\bar{g}^{\sigma\alpha}\xi^\beta\right)~, \label{DeltaH}
	\end{equation}
	which represents the Hamiltonian 2-form relative to the background spacetime.
	
	Once this relative 2-form is established, the quasi-local energy and angular momentum are obtained by integrating $\Delta H_{\mu\nu}$ over a closed spatial 2-surface $\mathcal{S}$. Assuming the spacetime admits isometries generated by a timelike Killing vector $t^\mu$ and an axial Killing vector $\varphi^\mu$, we define these relative charges as 
	\begin{align}
		\Delta_{\bar{g}}\mathcal{E}&\equiv\int_{\mathcal{S}}\Delta\mathbf{H}[t]=-\frac{1}{8\pi}\int_{\mathcal{S}}\left(\nabla^{[\alpha}t^{\beta]}+\nabla_\sigma\bar{g}^{\sigma[\alpha}t^{\beta]}\right)\mathrm{d}S_{\alpha\beta}~, \label{DeltaE} \\
		\Delta_{\bar{g}}\mathcal{J}&\equiv-\int_{\mathcal{S}}\Delta\mathbf{H}[\varphi]=\frac{1}{8\pi}\int_{\mathcal{S}}\left(\nabla^{[\alpha}\varphi^{\beta]}+\nabla_\sigma\bar{g}^{\sigma[\alpha}\varphi^{\beta]}\right)\mathrm{d}S_{\alpha\beta}~, \label{DeltaJ}
	\end{align}
	where the integration measure over the 2-surface $\mathcal{S}$ is given by $\mathrm{d}S_{\alpha\beta}=n_{\alpha\beta}\,\mathrm{d}u\mathrm{d}v$, with the binormal tensor $n_{\alpha\beta}$ defined in terms of the intrinsic coordinates $(u,v)$ of $\mathcal{S}$ as 
	\begin{equation}
		n_{\alpha\beta}\equiv\frac{1}{2}\epsilon_{\alpha\beta\mu\nu}\frac{\partial x^\mu}{\partial u}\frac{\partial x^\nu}{\partial v}~.
	\end{equation}
	
	Equations~\eqref{DeltaE} and \eqref{DeltaJ} represent the main results of this work, providing a concrete method to calculate quasi-local energy and angular momentum compatible with the Iyer-Wald formalism. Due to the matter contribution, these quasi-local quantities depend, in general, on the choice of the integration surface. To support this theoretical interpretation and demonstrate its physical viability, it is necessary to evaluate these expressions in concrete scenarios. In the following sections, we will apply this formalism to several cases of physical interest, showing how these relative charges encode the mass and angular momentum of specific black holes.
	
	\section{Applications on a Minkowski background}
	\label{sec:applicationsMin}
	
	\subsection{Spherically symmetric Kerr-Schild}
	
	\subsubsection{General setup and Misner-Sharp mass}
	
	It is well established that arbitrary spherically symmetric black-hole horizons, particularly those formalized as future outer trapping horizons, admit a quasi-local formulation of their thermodynamics based on the Misner-Sharp mass~\cite{hayward1994general, hayward1996gravitational, hayward2009local, padmanabhan2002classical, faraoni2015cosmological}, which is defined as a function of the areal radius $r$ by
	\begin{equation}
		M_{\text{MS}}(r)\equiv\frac{r}{2}\left(1-g^{\mu\nu}\partial_\mu r\partial_\nu r\right)~. \label{MS_definition}
	\end{equation}
	Motivated by this framework, we begin by establishing our formalism for a generic static, spherically symmetric metric that can be cast in the GKS form over a flat Minkowski background, $\bar{g}_{\mu\nu}=\eta_{\mu\nu}$, which is the traditional KS metric. In KS coordinates, the metric is $g_{\mu\nu}=\eta_{\mu\nu}+fk_\mu k_\nu$. We consider a generic KS scalar profile $f(r)$ and the radial null geodesic 1-form $k=\mathrm{d}t+\mathrm{d}r$. The spacetime admits an exact timelike Killing vector field $t^\mu=(\partial_t)^\mu$.
	
	For a KS metric, the inverse metric component in the radial direction is given by $g^{rr}=\eta^{rr}-f(k^r)^2=1-f$. Substituting this into Eq.~\eqref{MS_definition}, the Misner-Sharp mass for a generic spherical KS geometry reduces exactly to 
	\begin{equation}
		M_{\text{MS}}(r)=\frac{r}{2}f(r)~. \label{ms}
	\end{equation}
	
	Now, let us evaluate our quasi-local energy $\Delta_{\text{Mink}}\mathcal{E}$ for this geometry. Applying the integral \eqref{DeltaE} over a closed 2-sphere $\mathcal{S}$ of radius $r$, we obtain
	\begin{equation}
		\Delta_{\text{Mink}}\mathcal{E}_{\text{sph}}=\frac{r}{2}f(r)=M_{\text{MS}}~. \label{DeltaMinkE=MS}
	\end{equation}
	
	This proves that, in the spherical case, our quasi-local Hamiltonian recovers the Misner-Sharp mass. Since our formalism in Eq.~\eqref{DeltaE} relies only on the KS structure rather than on spherical symmetry, it remains well-defined when rotation is introduced. Consequently, our quasi-local Hamiltonian serves as a generalization of the Misner-Sharp mass for rotating black holes admitting a KS representation. Before considering this rotating case, to demonstrate the physical implications of this result, we apply it below to two fundamental spherically symmetric black hole solutions.
	
	\subsubsection{The Schwarzschild black hole}
	
	The simplest exact solution is the Schwarzschild black hole, whose KS scalar profile is 
	\begin{equation}
		f_{\text{Sch}}(r)=\frac{2M}{r}~.
	\end{equation}
	Substituting this function into our established equivalence~\eqref{DeltaMinkE=MS} immediately yields the correct global ADM mass at any radius $r$: 
	\begin{equation}
		\Delta_{\text{Mink}}\mathcal{E}_{\text{Sch}}=M~.
	\end{equation}
	To understand how our formalism achieves this, it is instructive to explicitly unpack the integral~\eqref{DeltaE} for this metric. The first part involves the standard Noether charge. For the Schwarzschild geometry, the evaluation yields the well-known result of half the Komar mass: 
	\begin{equation}
		-\frac{1}{8\pi}\int_{\mathcal{S}}\nabla^{[\alpha}t^{\beta]}\mathrm{d}S_{\alpha\beta}=\frac{M}{2}~. \label{half_komar}
	\end{equation}
	However, the second part of our quasi-local Hamiltonian~\eqref{DeltaE} requires the evaluation of the background term. Since $\bar{g}^{\mu\nu}=\eta^{\mu\nu}$ and the covariant derivative $\nabla$ is compatible with the full metric $g_{\mu\nu}$, one finds that this relative term evaluates to 
	\begin{equation}
		-\frac{1}{8\pi}\int_{\mathcal{S}}\nabla_\sigma\eta^{\sigma[\alpha}t^{\beta]}\mathrm{d}S_{\alpha\beta}=\frac{M}{2}~. \label{missing_half}
	\end{equation}
	Summing the contributions from Eqs.~\eqref{half_komar} and \eqref{missing_half}, the background reference term is explicitly responsible for supplying the missing half of the Komar mass, restoring the quasi-local energy to the correct value $M$.
	
	\subsubsection{The Reissner-Nordstr\"{o}m black hole}
	
	To investigate the physical distinction between this quasi-local energy and the global ADM mass, we consider the Reissner-Nordstr\"{o}m (RN) black hole. For an RN spacetime, the KS scalar is given by 
	\begin{equation}
		f_{\text{RN}}(r)=\frac{2M}{r}-\frac{Q^2}{r^2}~,
	\end{equation}
	where $Q$ is the electric charge. Substituting this profile into Eq.~\eqref{DeltaMinkE=MS}, the quasi-local energy evaluates to 
	\begin{equation}
		\Delta_{\text{Mink}}\mathcal{E}_{\text{RN}}(r)=M-\frac{Q^2}{2r}~. \label{quasi_local_RN}
	\end{equation}
	This result highlights the quasi-local nature of our formalism. At any finite radius $r$, the evaluated energy is less than the total ADM mass $M$. The subtracted term, $-Q^2/(2r)$, corresponds precisely to the classical electrostatic energy of the electric field stored in the exterior region (from the boundary $r$ to spatial infinity). By evaluating this quasi-local energy at the horizon, a thermodynamic picture emerges \cite{hayward1998unified}. The near-horizon electromagnetic stress generates a local thermodynamic pressure $P$. Tied to the quasi-local nature of the energy, this yields a first law of the form $\mathrm{d}M_{\text{MS}}=T\mathrm{d}S-P\mathrm{d}V$, where $T$ is the Hawking temperature, $S$ is the Bekenstein-Hawking entropy, and $V$ is the Euclidean volume enclosed by the horizon.
	
	\subsection{Axisymmetric Kerr-Schild}
	
	\subsubsection{General setup and generalized Misner-Sharp mass}
	
	We now extend the formalism to stationary, axisymmetric spacetimes. The Kerr-Schild metric takes the form $g_{\mu\nu}=\eta_{\mu\nu}+fk_\mu k_\nu$, with a generic scalar profile $f(r, \theta)$. With the Minkowski background written in oblate spheroidal coordinates $(t, r, \theta, \phi)$ \cite{11}, rotation is geometrically encoded in the null geodesic 1-form $k$, which acquires an azimuthal component $k_\phi=-a\sin^2\theta$, where $a$ is the angular momentum parameter. The spacetime admits two Killing vector fields, associated with time translation $t^\mu=(\partial_t)^\mu$ and axial rotation $\varphi^\mu=(\partial_\phi)^\mu$. Explicitly, the Minkowski background metric takes the form
	\begin{equation}
		\mathrm{d}\bar{s}^2=-\mathrm{d}t^2+\frac{\rho^2}{r^2+a^2}\mathrm{d}r^2+\rho^2 \mathrm{d}\theta^2+(r^2+a^2)\sin^2\theta \mathrm{d}\phi^2~,
	\end{equation}
	with $\rho^2=r^2+a^2\cos^2\theta$, where $a$ is the angular momentum parameter, and $k$ is given by
	\begin{equation}
		k_\mu \mathrm{d}x^\mu=\mathrm{d}t+\frac{\rho^2}{r^2+a^2}\mathrm{d}r-a\sin^2\theta \mathrm{d}\phi~.
	\end{equation}
	
	Applying our quasi-local energy formula~\eqref{DeltaE} over a closed 2-surface $\mathcal{S}$ of constant $t$ and $r$, the evaluation of the full antisymmetric tensor yields a boundary flux dependent on both $f(r,\theta)$ and its radial derivative. Integrating over the azimuthal angle $\phi\in[0,2\pi]$, the quasi-local energy reduces to 
	\begin{equation}
		\Delta_{\text{Mink}}\mathcal{E}_{\text{rot}}=\frac{1}{8}\int_0^\pi\left[2rf(r,\theta)-a^2\sin^2\theta\,\partial_r f(r,\theta)\right]\sin\theta\,\mathrm{d}\theta~. \label{E_rot_generic}
	\end{equation}
	Similarly, the quasi-local angular momentum associated with the axial Killing vector $\varphi^\mu$ via Eq.~\eqref{DeltaJ} yields 
	\begin{equation}
		\Delta_{\text{Mink}}\mathcal{J}_{\text{rot}}=\frac{a}{8}\int_0^\pi\left[2rf(r,\theta)-(r^2+a^2)\partial_r f(r,\theta)\right]\sin^3\theta\,\mathrm{d}\theta~. \label{J_rot_generic}
	\end{equation}
	
	Eqs.~\eqref{E_rot_generic} and \eqref{J_rot_generic} represent the generalized quasi-local charges for any generic rotating KS geometry. Before specializing to explicit black hole solutions, we verify the consistency of these generic formulas in the non-rotating limit. Taking $a\to 0$, the scalar profile recovers spherical symmetry, $f(r,\theta)\to f(r)$, and the angular derivative terms vanish. Evaluating Eq.~\eqref{E_rot_generic} under these conditions,
	\begin{equation}
		\lim_{a\to 0}\Delta_{\text{Mink}}\mathcal{E}_{\text{rot}}=\frac{1}{8}\int_0^\pi 2rf(r)\sin\theta\,\mathrm{d}\theta=\frac{r}{2}f(r)~.
	\end{equation}
	This result recovers Eq.~\eqref{DeltaMinkE=MS}, demonstrating that for any stationary KS spacetime, our quasi-local energy reduces to the spherical Misner-Sharp mass when rotation is removed. Similarly, the quasi-local angular momentum in Eq.~\eqref{J_rot_generic} vanishes.
	
	Several physically relevant stationary, axisymmetric black hole spacetimes are described by a G\"{u}rses-G\"{u}rsey metric~\cite{gurses1975lorentz}. For this family, the Kerr-Schild scalar is tied to a radial mass function $M(r)$, taking the form 
	\begin{equation}
		f(r,\theta)=\frac{2M(r)r}{r^2+a^2\cos^2\theta}~. \label{GG_profile}
	\end{equation}
	
	Substituting this structure into our generic integral formulas~\eqref{E_rot_generic} and~\eqref{J_rot_generic}, the boundary fluxes can be evaluated analytically. The integration over the polar angle yields the quasi-local charges of the G\"{u}rses-G\"{u}rsey metric: 
	\begin{align}
		& \Delta_{\text{Mink}}\mathcal{E}_{\text{GG}}=M(r)+\frac{ar-(r^2+a^2)\arctan(a/r)}{2a}\,M'(r)~, \label{E_master} \\
		& \Delta_{\text{Mink}}\mathcal{J}_{\text{GG}}=aM(r)-\frac{(r^2+a^2)\left[-ar+(r^2+a^2)\arctan(a/r)\right]}{2a^2}\,M'(r)~, \label{J_master}
	\end{align}
	where prime represents the derivative with respect to $r$. The energy $\Delta_{\text{Mink}}\mathcal{E}_{\text{GG}}$ decomposes into the Komar and background integral contributions, respectively given by
	\begin{align}
		&-\frac{1}{8\pi}\int_{\mathcal{S}}\nabla^{[\alpha}t^{\beta]}\mathrm{d}S_{\alpha\beta}=\frac{aM(r)-(a^2+r^2)\arctan(a/r)M'(r)}{2a}~, \label{Komar_Mink} \\
		&-\frac{1}{8\pi}\int_{\mathcal{S}}\nabla_\sigma\eta^{\sigma[\alpha}t^{\beta]}\mathrm{d}S_{\alpha\beta}=\frac{1}{2}\left[M(r)+rM'(r)\right]~. \label{Background_Mink}
	\end{align}
	These results capture the physics of frame-dragging. If the spacetime contains external matter or electromagnetic fields ($M'(r)\neq 0$), the rotation of the black hole couples with this external energy density, generating off-diagonal flux contributions.
	
	In light of the static limit $a\to 0$ and their geometric construction, the quasi-local energy expressions [both the generic integral in Eq.~\eqref{E_rot_generic} and the exact G\"{u}rses-G\"{u}rsey form in Eq.~\eqref{E_master})] can be understood as natural generalizations of the Misner-Sharp mass for rotating geometries. Analogously, the corresponding expressions for $\Delta_{\text{Mink}}\mathcal{J}$ in Eqs.~\eqref{J_rot_generic} and \eqref{J_master} serve as a complementary generalization, establishing a well-defined ``Misner-Sharp-like'' quasi-local angular momentum for axisymmetric Kerr-Schild spacetimes.
	
	\subsubsection{The Kerr-Newman black hole}
	
	For Kerr-Newman, the mass function is $M(r)=M-Q^2/(2r)$, with a term that accounts for the external electromagnetic energy density. Substituting into Eqs.~\eqref{E_master} and~\eqref{J_master}, the total quasi-local energy evaluates to 
	\begin{equation}
		\Delta_{\text{Mink}}\mathcal{E}_{\text{KN}}=M-\frac{Q^2}{4r}-\frac{Q^2(r^2+a^2)}{4ar^2}\arctan\left(\frac{a}{r}\right)~. \label{E_KN_total}
	\end{equation}
	The decomposition of Eq.~\eqref{E_KN_total} provides insight into the role of the background. From Eq.~\eqref{Komar_Mink}, the (Komar-like) Noether flux captures all electromagnetic and frame-dragging effects: 
	\begin{equation}
		-\frac{1}{8\pi}\int_{\mathcal{S}}\nabla^{[\alpha}t^{\beta]}\mathrm{d}S_{\alpha\beta}=\frac{M}{2}-\frac{Q^2}{4r}-\frac{Q^2(r^2+a^2)}{4ar^2}\arctan\left(\frac{a}{r}\right)~. \label{Noether_KN}
	\end{equation}
	In contrast, from Eq.~\eqref{Background_Mink}, the background contribution is
	\begin{equation}
		-\frac{1}{8\pi}\int_{\mathcal{S}}\nabla_\sigma\eta^{\sigma[\alpha}t^{\beta]}\mathrm{d}S_{\alpha\beta}=\frac{M}{2}~. \label{back_KN}
	\end{equation}
	This shows that the Minkowski reference serves exclusively, in this case, to restore the missing half of the mass parameter $M$.
	
	The quasi-local angular momentum $\Delta_{\text{Mink}}\mathcal{J}_{\text{KN}}(r)$ is found to be 
	\begin{equation}
		\Delta_{\text{Mink}}\mathcal{J}_{\text{KN}}=aM+\frac{Q^2(r^2-a^2)}{4ar}-\frac{Q^2(r^2+a^2)^2}{4a^2 r^2}\arctan\left(\frac{a}{r}\right)~.
	\end{equation}
	The $Q$-dependent terms in $\Delta_{\text{Mink}}\mathcal{J}_{\text{KN}}$ account for the angular momentum stored in the electromagnetic field.
	
	Summing both parts, Eqs.~\eqref{Noether_KN} and~\eqref{back_KN}, yields the total energy in Eq.~\eqref{E_KN_total}. This expression is in perfect agreement with the results obtained by Aguirregabiria \textit{et al.} via traditional pseudo-tensor descriptions~\cite{Aguirregabiria:1996}. However, our formalism arrives at this result through a covariant quasi-local Hamiltonian, rooted in the Iyer-Wald framework.
	
	In the non-rotating limit ($a\to 0$), the angular momentum vanishes identically, and the quasi-local energy reduces to the Reissner-Nordstr\"{o}m Misner-Sharp mass of Eq.~\eqref{quasi_local_RN}. Thus, the expressions for $\Delta_{\text{Mink}}\mathcal{E}_{\text{KN}}$ and $\Delta_{\text{Mink}}\mathcal{J}_{\text{KN}}$ derived here act as natural generalizations of these concepts for charged, rotating geometries. Furthermore, at spatial infinity ($r\to\infty$), we recover the global ADM parameters $M$ and $J=aM$. This consistency corroborates that our formalism provides a robust description of a quasi-local Hamiltonian.
	
	\section{Applications on an Anti-de Sitter background}
	\label{sec:applicationsAdS}
	
	\subsection{Spherically symmetric Kerr-Schild in AdS}
	
	\subsubsection{General setup and AdS regularization}
	
	We now apply our quasi-local formalism to spacetimes with a negative cosmological constant, $\Lambda=-3/l^2$, where $l$ is the AdS radius. For a generic static and spherically symmetric metric, the appropriate reference is the anti-de Sitter (AdS) background, $\bar{g}_{\mu\nu}$. The full spacetime can be cast in the GKS form $g_{\mu\nu}=\bar{g}_{\mu\nu}+fk_\mu k_\nu$. Here, $f(r)$ is the generic scalar profile, and, in usual spherical coordinates $(t,r,\theta,\phi)$, the radial null geodesic 1-form is $k=\mathrm{d}t+(1+r^2/l^2)^{-1}\mathrm{d}r$.
	
	We evaluate our quasi-local Hamiltonian~\eqref{DeltaE} over a closed 2-sphere $\mathcal{S}$ of radius $r$. The Komar-like integral yields 
	\begin{equation}
		-\frac{1}{8\pi}\int_{\mathcal{S}}\nabla^{[\alpha}t^{\beta]}\mathrm{d}S_{\alpha\beta}=\frac{r^3}{2l^2}-\frac{r^2}{4}f'(r)~.
	\end{equation}
	The background reference term is 
	\begin{equation}
		-\frac{1}{8\pi}\int_{\mathcal{S}}\nabla_\sigma\bar{g}^{\sigma[\alpha}t^{\beta]}\mathrm{d}S_{\alpha\beta}=\frac{r}{2}f(r)+\frac{r^2}{4}f'(r)~.
	\end{equation}
	Summing these contributions, the derivative terms cancel out, and the quasi-local energy, measured relative to the AdS background, is 
	\begin{equation}
		\Delta_{\text{AdS}}\mathcal{E}_{\text{sph}}(r) = \frac{r}{2}f(r)+\frac{r^3}{2l^2}~. 
		\label{E_sph_AdS}
	\end{equation}
	
	This reveals that evaluating the charge relative to the curved background captures the quasi-local energy $rf(r)/2$ of Eq.~\eqref{ms}, but inherently carries a boundary volume term $r^3/(2l^2)$ that diverges as $r\to\infty$. To evaluate the quasi-local energy relative to Minkowski, we add the energy measured with respect to the AdS background to the energy of AdS evaluated relative to flat space. This geometric decomposition acts as a background regularization procedure, defining a regularized energy $\mathcal{E}_{\text{sph}}$ via the additive prescription:
	\begin{equation}
		\mathcal{E}_{\text{sph}} = \Delta_{\text{AdS}}\mathcal{E}_{\text{sph}}+\Delta_{\text{Mink}}\mathcal{E}_{\text{AdS}}~.
	\end{equation}
	
	The energy of pure AdS measured relative to Minkowski is 
	\begin{equation}
		\Delta_{\text{Mink}}\mathcal{E}_{\text{AdS}}=-\frac{r^3}{2l^2}~. 
		\label{Delta_MinkE_AdS}
	\end{equation}
	By summing these contributions, the divergent background term is canceled: 
	\begin{equation}
		\mathcal{E}_{\text{sph}}=\left(\frac{r}{2}f(r)+\frac{r^3}{2l^2}\right)-\frac{r^3}{2l^2}=\frac{r}{2}f(r)~.
	\end{equation}
	The surviving source term, $rf(r)/2$, corresponds to the Misner-Sharp mass \eqref{ms} of the same black hole profile in a Minkowski background. Our additive formalism thus successfully renormalizes the asymptotic divergence of the quasi-local energy in AdS.
	
	\subsubsection{The Schwarzschild-AdS black hole}
	
	For the Schwarzschild-AdS solution, the GKS scalar profile is 
	\begin{equation}
		f_{\text{SAdS}}(r)=\frac{2M}{r}~.
	\end{equation}
	Once again, to understand the quasi-local charges in an AdS background, it is instructive to evaluate each term of Eq.~\eqref{DeltaE} separately. Substituting $f_{\text{SAdS}}$ and its derivative into our generic AdS evaluations, the (Komar-like) Noether term becomes
	\begin{equation}
		-\frac{1}{8\pi}\int_{\mathcal{S}}\nabla^{[\alpha}t^{\beta]}\mathrm{d}S_{\alpha\beta}=\frac{M}{2}+\frac{r^3}{2l^2}~.
	\end{equation}
	The background reference term is 
	\begin{equation}
		-\frac{1}{8\pi}\int_{\mathcal{S}}\nabla_\sigma\bar{g}^{\sigma[\alpha}t^{\beta]}\mathrm{d}S_{\alpha\beta}=\frac{M}{2}~.
	\end{equation}
	Just as in the Minkowski case, the background contributes the missing half of the mass, while remaining completely independent of the cosmological radius $l$. Summing these terms gives the total quasi-local energy relative to AdS: 
	\begin{equation}
		\Delta_{\text{AdS}}\mathcal{E}_{\text{SAdS}}(r)=M+\frac{r^3}{2l^2}~.
	\end{equation}
	
	Following our regularization procedure, adding the pure AdS vacuum energy~\eqref{Delta_MinkE_AdS}, the regularized mass 
	\begin{equation}
		\mathcal{E}_\text{SAdS}=M
	\end{equation}
	is recovered, decoupling the global mass parameter of the black hole from the diverging cosmological energy encapsulated by the boundary $\mathcal{S}$.
	
	\subsubsection{The Reissner-Nordstr\"{o}m-AdS black hole}
	
	When an electrostatic field is introduced, the GKS scalar profile for the Reissner-Nordstr\"{o}m-AdS spacetime takes the form 
	\begin{equation}
		f_{\text{RNAdS}}(r)=\frac{2M}{r}-\frac{Q^2}{r^2}~,
	\end{equation}
	where $Q$ represents the electric charge parameter. As with the previous cases, decomposing the boundary fluxes provides a clearer picture of the quasi-local Hamiltonian. Evaluating the pure Komar-like term with this profile yields 
	\begin{equation}
		-\frac{1}{8\pi} \int_{\mathcal{S}}\nabla^{[\alpha}t^{\beta]}\mathrm{d}S_{\alpha\beta}=\frac{M}{2}-\frac{Q^2}{2r}+\frac{r^3}{2l^2}~.
	\end{equation}
	This term captures the cosmological volume divergence, the exterior electrostatic energy, and only half of the mass parameter. The background reference contribution evaluates to 
	\begin{equation}
		-\frac{1}{8\pi} \int_{\mathcal{S}}\nabla_\sigma\bar{g}^{\sigma[\alpha}t^{\beta]} \mathrm{d}S_{\alpha\beta}=\frac{M}{2}~,
	\end{equation}
	resolving the standard half-mass anomaly, while remaining independent of both the electric charge $Q$ and the cosmological radius $l$.
	
	Summing these components, the total quasi-local energy relative to the AdS background becomes 
	\begin{equation}
		\Delta_{\text{AdS}}\mathcal{E}_{\text{RNAdS}}(r)=M-\frac{Q^2}{2r}+\frac{r^3}{2l^2}~.
	\end{equation}
	Applying our regularization procedure, the addition of the pure AdS energy~\eqref{Delta_MinkE_AdS} cancels the divergent volume term. The regularized quasi-local energy is thus recovered as 
	\begin{equation}
		\mathcal{E}_{\text{RNAdS}} =M-\frac{Q^2}{2r}~.
	\end{equation}
	
	This formulation correctly decouples the mass parameter $M$ and the energy contribution of the electrostatic field from the cosmological background. This regularized expression matches the Misner-Sharp mass of the asymptotically flat Reissner-Nordstr\"{o}m spacetime established in Eq.~\eqref{quasi_local_RN}, ensuring consistency across the varied geometric backgrounds.
	
	\subsection{Axisymmetric Kerr-Schild in AdS}
	
	\subsubsection{General setup and generalized AdS regularization}
	
	We now extend the formalism to stationary, axisymmetric spacetimes with a negative cosmological constant, \mbox{$\Lambda=-3/l^2$}. The GKS metric takes the form $g_{\mu\nu}=\bar{g}_{\mu\nu}+fk_\mu k_\nu$, with a generic scalar profile $f(r, \theta)$. With the AdS background $\bar{g}_{\mu\nu}$ written in the co-rotating Gibbons spheroidal coordinates $(t, r, \theta, \phi)$~\cite{11}, rotation is geometrically encoded in the null geodesic 1-form $k$. The spacetime admits two exact Killing vector fields associated with time translation $t^\mu=(\partial_t)^\mu$ and axial rotation $\varphi^\mu=(\partial_\phi)^\mu$. Explicitly, the AdS background metric takes the form
	\begin{equation}
		\mathrm{d}\bar{s}^2=-\frac{\Delta_\theta}{\Xi}\left(1+\frac{r^2}{l^2}\right)\mathrm{d}t^2+\frac{\rho^2}{\Delta_r}\mathrm{d}r^2+\frac{\rho^2}{\Delta_\theta}\mathrm{d}\theta^2+\frac{(r^2+a^2)\sin^2\theta}{\Xi}\mathrm{d}\phi^2~, \label{Background_AdS_rot}
	\end{equation}
	with $\rho^2=r^2+a^2\cos^2\theta$, $\Xi=1-a^2/l^2$, $\Delta_\theta=1-a^2/l^2\,\cos^2\theta$, and $\Delta_r=(r^2+a^2)\left(1+r^2/l^2\right)$. The null geodesic 1-form $k$ is given by
	\begin{equation}
		k_\mu \mathrm{d}x^\mu = \frac{\Delta_\theta}{\Xi}\mathrm{d}t+\frac{\rho^2}{\Delta_r} \mathrm{d}r-\frac{a\sin^2\theta}{\Xi}\mathrm{d}\phi~.
	\end{equation}
	
	Applying our quasi-local energy formula~\eqref{DeltaE} over a closed 2-surface $\mathcal{S}$ of constant $t$ and $r$, and utilizing the co-rotating AdS background, the evaluation of the Hamiltonian 2-form yields a boundary flux dependent on the profile $f(r,\theta)$ and its radial derivative. Integrating over the azimuthal angle $\phi\in[0,2\pi]$, the exact quasi-local energy functional reduces to 
	\begin{equation}
		\Delta_{\text{AdS}}\mathcal{E}_{\text{rot}}=\frac{r(r^2+a^2)}{2l^2\Xi}+\frac{1}{8\Xi^2}\int_0^\pi\left[2r\Delta_\theta f(r,\theta)-a^2\sin^2\theta\left(1+\frac{r^2}{l^2}\right)\partial_r f(r,\theta)\right]\sin\theta\,\mathrm{d}\theta~.
		\label{E_rot_AdS_generic}
	\end{equation}
	This result captures the geometric deformations induced by the cosmological background.
	
	Similarly, the quasi-local angular momentum associated with the axial Killing vector $\varphi^\mu$ via Eq.~\eqref{DeltaJ} yields 
	\begin{equation}
		\Delta_{\text{AdS}}\mathcal{J}_{\text{rot}}=\frac{a}{8\Xi^2}\int_0^\pi\left[2rf(r,\theta)-(r^2+a^2)\partial_r f(r,\theta)\right]\sin^3\theta\,\mathrm{d}\theta~. \label{J_AdS_generic}
	\end{equation}
	Unlike the energy functional in Eq.~\eqref{E_rot_AdS_generic}, the AdS radius $l$ does not introduce angular or radial corrections to the integrand itself. Instead, its entire contribution factors out as the scaling factor $\Xi^{-2}$.
	
	The rotating AdS background metric given by Eq.~\eqref{Background_AdS_rot} can itself be expressed over a flat Minkowski spacetime in a KS form. In Gibbons spheroidal coordinates, this Kerr-Schild decomposition reads $\bar{g}_{\mu\nu}=\eta_{\mu\nu}+f_{\text{AdS}}k^{\text{AdS}}_\mu k^{\text{AdS}}_\nu$, where the Minkowski background metric $\eta_{\mu\nu}$ is
	\begin{align}
		\mathrm{d}s^2_{\text{Mink}} &=-\mathrm{d}t^2+\frac{(r^2+a^2)\sin^2\theta}{\Xi}\mathrm{d}\phi^2+\left[\frac{\rho^2}{\Delta_r}+\frac{r^2}{l^2(1+r^2/l^2)^2}\right]\mathrm{d}r^2+\left[\frac{\rho^2}{\Delta_\theta}+\frac{a^4\sin^2\theta\cos^2\theta}{l^2\Delta_\theta^2}\right]\mathrm{d}\theta^2 \notag \\
		& \quad+\frac{2rR}{l^2(1+r^2/l^2)}\mathrm{d}t\mathrm{d}r+\frac{2a^2R\sin\theta\cos\theta}{l^2\Delta_\theta}\mathrm{d}t\mathrm{d}\theta+\frac{2a^2r\sin\theta\cos\theta}{l^2(1+r^2/l^2)\Delta_\theta} \mathrm{d}r\mathrm{d}\theta~.
	\end{align}
	Introducing $R^2=(r^2\Delta_\theta+a^2\sin^2\theta)/\Xi$, the KS scalar profile is given by $f_{\text{AdS}}=-R^2/l^2$. The associated null geodesic 1-form $k^\text{AdS}$ is
	\begin{equation}
		k^{\text{AdS}}_\mu \mathrm{d}x^\mu=\mathrm{d}t+\frac{r}{(1+r^2/l^2)R}\mathrm{d}r+\frac{a^2\sin\theta\cos\theta}{\Delta_\theta R}\mathrm{d}\theta~.
	\end{equation}
	Evaluating the AdS background relative to this flat Minkowski reference, we find that the relevant component of the Hamiltonian 2-form is
	\begin{equation}
		\Delta_{\text{Mink}}H_{\text{AdS}}^{tr} = -\frac{r(r^2+a^2)}{8\pi l^2(r^2+a^2\cos^2\theta)}~. \label{Htr_Mink}
	\end{equation}
	
	Integrating over the closed 2-surface $\mathcal{S}$, the corresponding quasi-local energy of the pure AdS background measured relative to Minkowski is found to be 
	\begin{equation}
		\Delta_{\text{Mink}}\mathcal{E}_{\text{rotAdS}}=\int_{\mathcal{S}}\Delta_{\text{Mink}}H_{\text{AdS}}^{\mu\nu}\,\mathrm{d}S_{\mu\nu}=-\frac{r(r^2+a^2)}{2l^2\Xi}~.
		\label{E_AdS_Mink}
	\end{equation}
	This provides a natural regularization procedure for the general Kerr-Schild geometries on an AdS background. The regularized energy relative to Minkowski, denoted simply as $\mathcal{E}_{\text{rot}}$, takes the form
	\begin{equation}
		\mathcal{E}_{\text{rot}} = \Delta_{\text{AdS}}\mathcal{E}_{\text{rot}}+\Delta_{\text{Mink}}\mathcal{E}_{\text{rotAdS}} = \frac{1}{8\Xi^2}\int_0^\pi\left[2r\Delta_\theta f(r,\theta) - a^2\sin^2\theta\left(1+\frac{r^2}{l^2}\right)\partial_r f(r,\theta)\right]\sin\theta\,\mathrm{d}\theta~. 
		\label{E_rot_Minkowski}
	\end{equation}
	
	A similar additive regularization procedure extends to the angular momentum. However, since the pure AdS vacuum has no intrinsic rotation, its quasi-local angular momentum measured relative to the flat Minkowski background vanishes ($\Delta_{\text{Mink}}\mathcal{J}_{\text{AdS}}=0$). Consequently, the total regularized angular momentum for the general Kerr-Schild geometry on an AdS background, denoted simply as $\mathcal{J}_{\text{rot}}$, is entirely determined by its contribution relative to the AdS background: 
	\begin{equation}
		\mathcal{J}_{\text{rot}}=\Delta_\text{AdS}\mathcal{J}_\text{rot }+ \Delta_\text{Mink}\mathcal{J}_\text{AdS}=\Delta_\text{AdS}\mathcal{J}_\text{rot}~. \label{J_rot_Minkowski_add}
	\end{equation}
	
	As in the asymptotically flat case, several physically relevant stationary, axisymmetric black hole spacetimes are described by a generalized G\"{u}rses-G\"{u}rsey metric. In this case, the GKS scalar takes the form given in Eq.~\eqref{GG_profile}. Substituting this into Eqs.~\eqref{E_rot_AdS_generic} and~\eqref{J_AdS_generic}, the integrals are evaluated as 
	\begin{align}
		&\Delta_{\text{AdS}}\mathcal{E}_{\text{GG}}=\frac{r(r^2+a^2)}{2l^2\Xi}+\frac{M(r)}{\Xi^2}-\frac{1}{2a\Xi^2}\left(1+\frac{r^2}{l^2}\right)\left[(r^2+a^2)\arctan\left(\frac{a}{r}\right)-ar\right]M'(r)~, \label{E_GG_exact} \\
		& \Delta_{\text{AdS}}\mathcal{J}_{\text{GG}}=\frac{aM(r)}{\Xi^2}-\frac{r^2+a^2}{2a^2\Xi^2}\left[(r^2+a^2)\arctan\left(\frac{a}{r}\right)-ar\right]M'(r)~. \label{J_GG_exact}
	\end{align}
	It immediately follows from Eq.~\eqref{E_rot_Minkowski} that the regularized energy relative to Minkowski for the generalized G\"{u}rses-G\"{u}rsey metric is 
	\begin{equation}
		\mathcal{E}_{\text{GG}} =\frac{M(r)}{\Xi^2}-\frac{1}{2a\Xi^2}\left(1+\frac{r^2}{l^2}\right) \left[(r^2+a^2)\arctan\left(\frac{a}{r}\right)-ar\right]M'(r)~. \label{E_GG_Minkowski}
	\end{equation}
	
	\subsubsection{The Kerr-AdS black hole}
	
	For the uncharged, rotating Kerr-AdS black hole, the spacetime is a pure vacuum solution. Consequently, the mass profile of the generalized G\"{u}rses-G\"{u}rsey metric reduces to a constant, $M(r)=M$, representing the geometric mass parameter.
	
	Before applying the background regularization procedure, we evaluate the quasi-local charges measured relative to the AdS background. Substituting this profile into Eqs.~\eqref{E_GG_exact} and \eqref{J_GG_exact}, the derivative terms drop out completely, yielding the unregularized quasi-local energy and angular momentum: 
	\begin{equation}
		\Delta_{\text{AdS}} \mathcal{E}_{\text{KAdS}}=\frac{M}{\Xi^2}+\frac{r(r^2+a^2)}{2l^2\Xi}~,
		\qquad\Delta_{\text{AdS}}\mathcal{J}_{\text{KAdS}}=\frac{aM}{\Xi^2}~. \label{KerrAdS_unreg}
	\end{equation}
	As expected, while the angular momentum is inherently finite, the unregularized energy contains a boundary volume term that diverges as $r\to\infty$.
	
	Now, we invoke our regularization procedure by utilizing the additive Eqs.~\eqref{E_rot_Minkowski} and~\eqref{J_rot_Minkowski_add}. The divergent vacuum term is canceled by the AdS energy contribution relative to a Minkowski background~\eqref{E_AdS_Mink}, leaving only the globally conserved thermodynamic charges 
	\begin{equation}
		\mathcal{E}_{\text{KAdS}} =\frac{M}{\Xi^2}~,\qquad\mathcal{J}_{\text{KAdS}}= \frac{aM}{\Xi^2}~.
	\end{equation}
	
	This shows that our additive framework eliminates the divergent AdS vacuum contributions without relying on infinite boundary counterterms. By doing so, we recover the classical thermodynamic quantities of the Kerr-AdS black hole established by Gibbons \textit{et al.}~\cite{11} and Caldarelli \textit{et al.}~\cite{caldarelli}. These standard results constitute the framework referred to as the ``usual thermodynamic theory'' for Kerr-AdS in~\cite{campos2024generating,campos2025black}.
	
	\subsubsection{The Kerr-Newman-AdS black hole}
	
	When an electromagnetic field is introduced, the resulting Kerr-Newman-AdS spacetime is no longer a vacuum solution. The energy density of the exterior electrostatic field modifies the mass profile, which takes the specific form $M(r)=M-Q^2/(2r)$, where $Q$ is the electric charge parameter.
	
	As in the uncharged case, evaluating the energy relative to the AdS background yields an unregularized quasi-local energy containing a divergent vacuum boundary volume term. By applying our additive regularization framework, we obtain the regularized quasi-local energy for the Kerr-Newman-AdS black hole: 
	\begin{equation}
		\mathcal{E}_{\text{KNAdS}}=\frac{M}{\Xi^2}-\frac{Q^2}{2r\Xi^2}-\frac{Q^2}{4ar^2\Xi^2}\left(1+\frac{r^2}{l^2}\right)\left[(r^2+a^2)\arctan\left(\frac{a}{r}\right)-ar\right]~. \label{E_KNAdS}
	\end{equation}
	Similarly, substituting the profile into Eq.~\eqref{J_rot_Minkowski_add} yields the regularized quasi-local angular momentum: 
	\begin{equation}
		\mathcal{J}_{\text{KNAdS}}= \frac{aM}{\Xi^2}-\frac{aQ^2}{2r\Xi^2}-\frac{Q^2(r^2+a^2)}{4a^2 r^2\Xi^2} \left[(r^2+a^2)\arctan\left(\frac{a}{r}\right)-ar\right]~. \label{J_KNAdS}
	\end{equation}
	These equations capture the coupling between the black hole's rotation, its electromagnetic field, and the AdS geometry, with the leading terms representing globally conserved charges.
	
	In the non-rotating limit ($a\to 0$), the angular momentum vanishes, and the quasi-local energy reduces to the Reissner-Nordstr\"{o}m-AdS result, recovering the generalized Misner-Sharp mass for this geometry. Furthermore, taking the limit of spatial infinity ($r\to\infty$) isolates the standard thermodynamic quantities $M/\Xi^2$ and $aM/\Xi^2$. This consistency further confirms that our covariant Hamiltonian approach provides a coherent quasi-local formulation for general black-hole solutions of KS form.
	
	Eqs.~\eqref{E_KNAdS} and \eqref{J_KNAdS} can be regarded as a generalization of the results of Aguirregabiria \textit{et al.} \cite{Aguirregabiria:1996} to an AdS background. However, as in the pure Kerr-Newman case, rather than relying on pseudo-tensors, we arrive at this result through the completely covariant quasi-local Hamiltonian framework of the Iyer-Wald formalism.
	
	\subsection{Smarr formula in extended phase space}
	
	The robustness of our regularized quasi-local charges can be further demonstrated by showing that our quasi-local Hamiltonian formalism encodes the extended Smarr relation when evaluated at the event horizon. In the extended phase space, the cosmological constant is interpreted as a thermodynamic pressure, $P=-\Lambda/(8\pi)$, with conjugate variable given by the thermodynamic volume $V$.
	
	For a stationary rotating black hole, such as Kerr-AdS, the event horizon $\mathcal{H}$ (located at $r=r_+$) is generated by the null Killing vector field
	\begin{equation}
		\xi^\mu=t^\mu+\Omega_H\varphi^\mu~, \label{kv}
	\end{equation}
	where $\Omega_H$ is the angular velocity of the horizon. Due to the linearity of the Hamiltonian operator $\Delta_{\text{AdS}}\mathbf{H}$, the integration of the Hamiltonian 2-form associated with $\xi^\mu$ over the horizon cross-section splits into the contributions from time translation and axial rotation: 
	\begin{equation}
		\int_{\mathcal{H}}\Delta_{\text{AdS}}\mathbf{H}[\xi^\mu] = \int_{\mathcal{H}}\Delta_{\text{AdS}}\mathbf{H}[t^\mu]+\Omega_H\int_{\mathcal{H}}\Delta_{\text{AdS}}\mathbf{H}[\varphi^\mu]~.  \label{H_linearity}
	\end{equation}
	
	The right-hand side of Eq.~\eqref{H_linearity} represents the unregularized quasi-local energy and angular momentum evaluated at the horizon. On the left-hand side, evaluating the Hamiltonian for the horizon-generating Killing vector $\xi^\mu$ yields the standard thermal Noether charge $\kappa A_H/(8\pi)=TS$ (where $\kappa$ is the surface gravity, $A_H$ is the area of $\mathcal{H}$, $T$ is the Hawking temperature, and $S$ is the entropy) alongside a background correction, $M/(2\Xi)$. Identifying the total global charges $\mathcal{E}$ and $\mathcal{J}$ through our regularization scheme of Eq.~\eqref{E_rot_Minkowski} ($\Delta_{\text{AdS}}\mathcal{E}=\mathcal{E}-\Delta_{\text{Mink}}\mathcal{E}_{\text{rotAdS}}$), the evaluation at the horizon takes the form 
	\begin{equation}
		TS +\frac{M}{2\Xi}=\mathcal{E}-\Omega_H\mathcal{J}-\Delta_{\text{Mink}} \mathcal{E}_{\text{AdS}}(r_+)~. 
		\label{Smarr_intermediate}
	\end{equation}
	
	Working in the traditional coordinate system for Kerr-AdS \cite{hawking1999rotation}, the Killing vectors are
	\begin{equation}
		t^\mu=(\partial_t)^\mu~,\qquad\varphi^\mu =(\partial_\phi)^\mu~, 
		\label{kvco}
	\end{equation}
	with the horizon angular velocity given by $\Omega_H=a\Xi/(r_+^2+a^2)$. In this frame, the global mass is related to the scaling factor by $\mathcal{E}=M/\Xi$. Consequently, Eq.~\eqref{Smarr_intermediate} reads
	\begin{equation}
		\frac{\mathcal{E}}{2} = TS+\Omega_H\mathcal{J}+\Delta_\text{Mink} \mathcal{E}_\text{AdS}(r_+)~.
	\end{equation}
	
	From Eq.~\eqref{E_AdS_Mink}, the last term on the right-hand side can be written as
	\begin{equation}
		\Delta_{\text{Mink}}\mathcal{E}_{\text{AdS}}(r_+)=-PV_{\text{geo}}~,\qquad V_{\text{geo}}=\frac{4\pi}{3}\frac{(r_+^2+a^2)r_+}{\Xi}~, \label{eads}
	\end{equation}
	where $V_{\text{geo}}$ is the vector volume generated by the Killing vector \eqref{kv} with~\eqref{kvco}. This quantity coincides with the geometric volume of the black hole, viewed as an ellipsoid of revolution in Euclidean coordinates \cite{ballik2013vector}. Hence, Eq.~\eqref{Smarr_intermediate} becomes
	\begin{equation}
		\mathcal{E} = 2TS +2\Omega_H\mathcal{J}-2PV_\text{geo}~.
	\end{equation}
	This represents a ``geometric'' Smarr formula since it is not an Euler relation, aligning with Hawking's original description of Kerr-AdS black holes~\cite{campos2024generating, hawking1999rotation}.
	
	Alternatively, we can transition to a frame co-rotating with the AdS boundary at infinity by defining the generators
	\begin{equation}
		t^\mu=(\partial_t)^\mu - \frac{a}{l^2}(\partial_\phi)^\mu~, \qquad \varphi^\mu=(\partial_\phi)^\mu~,
	\end{equation}
	where the angular velocity becomes $\Omega=\Omega_H+a/l^2$. In this co-rotating frame, we find $\mathcal{E}_\text{co}=M/\Xi^2$. Utilizing the geometric relation $\Xi=1+\Lambda\mathcal{J}^2/(3\mathcal{E}_\text{co}^2)$ to express $\Xi$ in terms of the globally conserved charges, Eq.~\eqref{Smarr_intermediate} becomes
	\begin{equation}
		\frac{\mathcal{E}_\text{co}}{2}=TS+\Omega\mathcal{J}+\Delta_{\text{Mink}}\mathcal{E}_{\text{AdS}}(r_+)+\frac{\Lambda\mathcal{J}^2}{6\mathcal{E}_\text{co}}~. \label{fl}
	\end{equation}
	Identifying the last two terms on the right-hand side of Eq.~\eqref{fl} as $-PV$, where $V$ is the thermodynamic volume conjugate to $P$, we recover the proper extended phase space Smarr formula for Kerr-AdS:
	\begin{equation}
		\mathcal{E}_\text{co} = 2TS+2\Omega_H\mathcal{J}-2PV~, \qquad V=V_{\text{geo}}+\frac{4\pi}{3}\frac{\mathcal{J}^2}{\mathcal{E}}~.
	\end{equation}
	This is a thermodynamic formulation, as it satisfies the Euler relation and is consistent with the standard extended phase space framework for Kerr-AdS black holes~\cite{11,caldarelli, campos2024generating}.
	
	This derivation reveals that the thermodynamic volume $V$ is intrinsically tied to the background reference energy $\Delta_{\text{Mink}}\mathcal{E}_{\text{AdS}}(r_+)$, and explains the origin of the difference between the thermodynamic and the geometric volumes. This proves that our geometric regularization procedure not only guarantees finite asymptotic charges, but also successfully embeds the mechanical pressure-volume dynamics into the quasi-local Hamiltonian structure of Kerr-AdS.
	
	\section{Final Remarks}
	\label{remarks}
	
	Extracting integrable charges in the Iyer-Wald formalism requires identifying a suitable boundary term whose variation matches the contribution of the symplectic potential in the Hamiltonian variation. The main objective of this paper is to explicitly derive this boundary term for GKS geometries, presenting the results in Eqs.~\eqref{DeltaE} and~\eqref{DeltaJ}. Consequently, by providing the Hamiltonian 2-form, our framework enables the evaluation of quasi-local Hamiltonians over finite domains for a broad class of spacetimes. This approach completely bypasses the need to add \textit{ad hoc} counterterms to the action, impose particular asymptotic boundary conditions, or perform an isometric embedding into a reference background.
	
	The general structure of our quasi-local Hamiltonian consists of a term that is completely determined by the Einstein-Hilbert Lagrangian, namely the Noether charge associated with the Komar integrals, plus a term that depends explicitly on the background metric. When applied to spherically symmetric KS spacetimes in a Minkowski background, our proposal reproduces the standard Misner-Sharp mass. Because our formalism relies exclusively on the Kerr-Schild form of the metric, it naturally provides a generalization of the Misner-Sharp mass for rotating (axisymmetric) geometries.
	
	Furthermore, extending our proposal to AdS backgrounds provides a quasi-local perspective on Kerr-AdS thermodynamics, successfully deducing the Smarr formula from purely geometric arguments. In our approach, the energy associated with the cosmological constant emerges when evaluating the AdS quasi-local Hamiltonian relative to a Minkowski reference. This contribution arises during the regularization process required to obtain the traditional globally conserved charges, successfully reproducing the results of extended phase-space thermodynamics established by previous works~\cite{dolan2011pressure,dolan2012pdv}. Our development explicitly uncovers the geometric reason behind the well-known discrepancy between the geometric volume and the thermodynamic volume of rotating AdS black holes.
	
	Beyond providing a closed-form expression for the quasi-local Hamiltonian of GKS black holes, this work takes a first step toward reconciling traditionally distinct approaches to black hole thermodynamics. Namely, it bridges quasi-local frameworks anchored on the Misner-Sharp mass with the covariant phase space framework of the Iyer-Wald formalism. Notable examples of these quasi-local approaches include the dynamical formulations by Hayward~\cite{hayward1994general,hayward1996gravitational,hayward1998unified} and the developments of Padmanabhan~\cite{padmanabhan2002classical}, where the gravitational field equations emerge as a direct consequence of a thermodynamic interpretation on the horizon. 
	
	By successfully embedding these thermodynamic perspectives within a variational formalism, our framework opens new avenues of research. The generalized Misner-Sharp mass derived here could serve as the foundation for an axisymmetric extension of the black hole thermodynamics proposed by Hayward for trapping horizons in the stationary limit. Furthermore, applying our quasi-local approach to the Euclidean action formalism may provide novel insights into the quantum statistical relation (QSR)~\cite{gibbons1977action,gibbons1978black}. Specifically, a formalism in which the Misner-Sharp mass acts as the energy of the black hole necessarily leads to an alternative QSR for various solutions, including electrically charged black holes. Finally, extending this procedure to higher-dimensional spacetimes, such as five-dimensional asymptotically AdS geometries, offers a promising route to explore the gauge/gravity duality, potentially establishing a holographic boundary dual for our quasi-local energy.
	
	\begin{acknowledgments}
		M. A. J. thanks Coordenação de Aperfeiçoamento de Pessoal de Nível Superior (CAPES) – Brazil, Finance Code 001, for financial support.
		T.~L.~C. acknowledges the support of Funda\c{c}\~ao Carlos Chagas Filho de Amparo \`{a} Pesquisa do Estado do Rio de Janeiro (FAPERJ) -- Brazil, Grant No.~E-26/202.775/2026 and 202.776/2026.
		M. C. B. thanks Thiago S. Pereira, from the State University of Londrina, for valuable discussions.
	\end{acknowledgments}

\end{document}